\documentclass[prl, twocolumn,showpacs,preprintnumbers,amsmath,amssymb]{revtex4}

\usepackage{graphicx}
\usepackage{bm}

\begin{document}
\preprint{}
\title{An energy scale directly related to superconductivity in the high-$T_c$ cuprate superconductors: Universality from the Fermi arc picture}
\author{S. Ideta$^{1*}$, T. Yoshida$^1$, A. Fujimori$^1$, H. Anzai$^2$, T. Fujita$^2$, A. Ino$^2$, M. Arita$^3$, H. Namatame$^3$, M. Taniguchi$^{2, 3}$, Z.-X. Shen$^4$, K. Takashima$^1$, K. Kojima$^1$, and S. Uchida$^1$}
\affiliation{
$^1$Department of Physics, University of Tokyo, Bunkyo-ku, Tokyo 113-0033, Japan\\
$^2$Graduate School of Science, Hiroshima University, Higashi-Hiroshima 739-8526, Japan\\
$^3$Hiroshima Synchrotron Center, Hiroshima University, Higashi-Hiroshima 739-0046, Japan\\
$^4$Department of Applied Physics and Stanford Synchrotron Radiation Laboratory, Stanford University, Stanford, CA94305
}

\date{\today}

\begin{abstract}
We have performed a temperature dependent angle-resolved photoemission spectroscopy (ARPES) study of the tri-layer high-$T_c$ cuprate superconductor (HTSC) Bi$_2$Sr$_2$Ca$_2$Cu$_3$O$_{10+\delta}$ (Bi2223), and have shown that the \textquotedblleft effective\textquotedblright superconducting (SC) gap $\Delta_{\rm{sc}}$ defined at the end point of the Fermi arc and the $T_c$ (= 110 K) approximately satisfies the weak-coupling BCS-relationship 2$\Delta_{\rm{sc}}$ = 4.3$k_{\rm{B}}T_c$. Combining this result with previous ARPES results on single- and double-layer cuprates, we show that the relationship between 2$\Delta_{\rm{sc}}$ = 4.3$k_{\rm{B}}T_c$ holds for various HTSCs. Furthermore, at $T$ $\sim$ $T_c$, the quasi-patricle width at the end point of the Fermi arc is found to coincide with $\Delta_{\rm{sc}}$, consistent with the context of Planckian dissipation.
\end{abstract}
\pacs{74.25.Jb, 71.18.+y, 74.72.Dn, 79.60.-i}

\maketitle
In the normal state of cuprate high-temperature superconductors (HTSCs), a pseudogap exists on part of the Fermi surface (FS) away from the $d$-wave superconducting (SC) gap node, and the FS is truncated into gapless regions called \textquotedblleft Fermi arcs\textquotedblright \cite{Kanigel, Yoshida, M. R. Norman, Kondo}. The most controversial and crucial questions are how the Fermi arc/pseudogap and the $d$-wave superconductivity are mutually related and how the $T_c$ is determined. Most remarkably, the superfluid density measured by muon spin relaxation ($\mu$SR) is proportional to $T_c$ for a large number of HTSCs, known as Uemura relation \cite{Uemura1}. It has also been found that the coherence peak in angle-resolved photoemission spectroscopy (ARPES) spectra around ($\pi$, 0) of Bi$_2$Sr$_2$CaCu$_2$O$_{8+\delta}$ (Bi2212) increases with the superfluid density and hence is proportional to $T_c$ \cite{Feng3}. As for the SC gap, while it had been believed from ARPES and scanning tunneling microscopy (STM) studies \cite{Lee, Yoshida, Kohsaka} that the gap increases with underdoping, the SC gap $\Delta_0$ around the node defined by the $d$-wave order parameter $\Delta(\overrightarrow{k})$ = $\Delta_0(\cos(k_xa)-\cos(k_ya))$/2, where $a$ is the lattice constant, estimated from the penetration depth and Andreev reflection measurements was found to be proportional to $T_c$ from the underdoped to slightly overdoped regions \cite{Panagopoulos, Deutscher}. Also, according to recent ARPES studies, $\Delta_0$ for La$_{2-x}$Sr$_x$CuO$_4$ (LSCO, $n$ = 1) and Bi2212 ($n$ = 2) ceases to increase with underdoping and then drop in non-superconducting samples \cite{Yoshida, Tanaka, Lee}. 

In order to understand the relationship between the $T_c$, the SC gap in the nodal region $\Delta_0$, and the superfluid density, a \textquotedblleft Fermi arc\textquotedblright picture has been proposed by  Lee and Wen \cite{LeePA} and Oda $et\ al$.\cite{Oda}. According to this picture, the length of the Fermi arc is proportional to the number of doped carriers $x$, leading to an \textquotedblleft effective SC gap\textquotedblright $\Delta_{\rm{sc}} \propto x\Delta_0$ defined by the gap magnitude at the edge of the Fermi arc, and hence $T_c$ is proportional to $x\Delta_0$. For further understanding of a \textquotedblleft{Fermi arc}\textquotedblright picture, it would be useful to investigate the psudogap and Fermi arc of tri-layer HTSC which has the highest $T_c$ among the cuprates. 

In this Letter, we report on a temperarure dependent ARPES results on the tri-layer HTSC Bi$_2$Sr$_2$Ca$_2$Cu$_3$O$_{10+\delta}$ (Bi2223) which shows the highest $T_c$ among the Bi-based HTSCs. From the present and previous ARPES results on various cuprates with different $T_c$'s, we have found that an \textquotedblleft effective\textquotedblright\ SC gap $\Delta_{\rm{sc}}$, which is defined by the gap at the end point of the Fermi arc, is proportional to $T_c$. This relationship takes an apparently weak-coupling BCS form 2$\Delta_{\rm{sc}} \simeq 4.3k_{\rm{B}}T_c$, and is found to be universal, providing an extension of the BCS relationship to the superconductivity of the HTSCs, where the carrier density is strongly reduced by electron correlation compared to conventional superconductors.

Single crystals of optimally doped Bi2223 ($T_c$ = 110 K) were grown by the travelling solvent floating zone (TSFZ) method. ARPES experiments were carried out at beam line 9A of Hiroshima Synchrotron Radiation Center (HiSOR) using circularly polarized light. The total energy resolution $\Delta E$ was set at $\sim$ 5 meV. The samples were cleaved $in\ situ$ under an ultrahigh vacuum of $\sim$ 1$\times$10$^{-11}$ Torr.

\begin{figure}[ht]
\includegraphics[width=8cm]{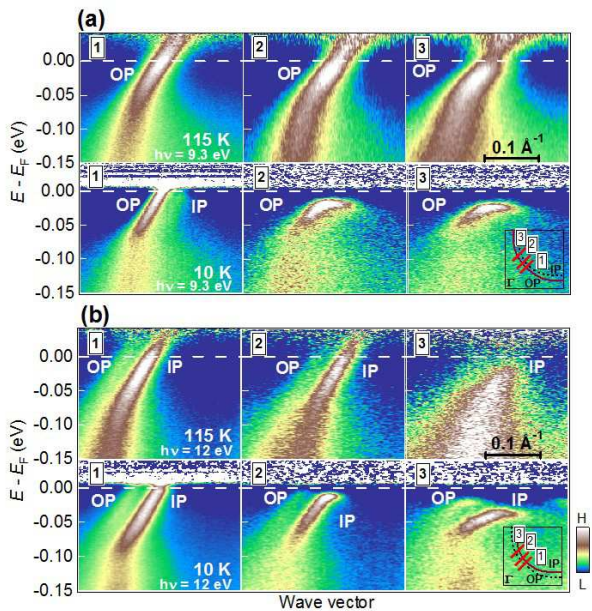}
\caption{(Color online) ARPES spectra of optimally doped Bi2223 ($T_c$ = 110 K) in the SC ($T$ = 10 K) and normal ($T$ = 115 K) states. (a, b): ARPES intensity plots divided by the Fermi-Dirac function convoluted with the resolusion function along the cuts shown by red lines in inset across the FS for the OP and IP bands, where OP and IP stand for the outer and inner CuO$_2$ planes, respectively. }
\end{figure}

Figure 1 shows an ARPES intensity plots in energy-momentum space of optimally doped Bi2223 ($T_c$ = 110 K) at $T$ = 10, 115 K for the nodal and the off-nodal cuts, demonstrating the temperature and momentum dependences of the energy gap. Here, by utilizing the photon-energy dependence of ARPES intensities, one can observe the outer and inner CuO$_2$ planes (OP and IP) separately \cite{Ideta}. In the SC state ($T$ = 10 K), the gap opens except for the nodal direction as expected for a $d$-wave superconductor. In the normal state ($T$ = 115 K), slightly above $T_c$, there is a gapless region around the nodal direction [1$-$2 of panel (a) at 115 K and 1 of panel (b) at 115 K] by which we define the \textquotedblleft Fermi arc\textquotedblright region. Near the anti-nodal direction, the pseudogap remains open at 115 K as in 3 of panel (a) and 2$-$3 of panel (b) in Fig. 1. 
\begin{figure*}[ht]
\includegraphics[width=16cm]{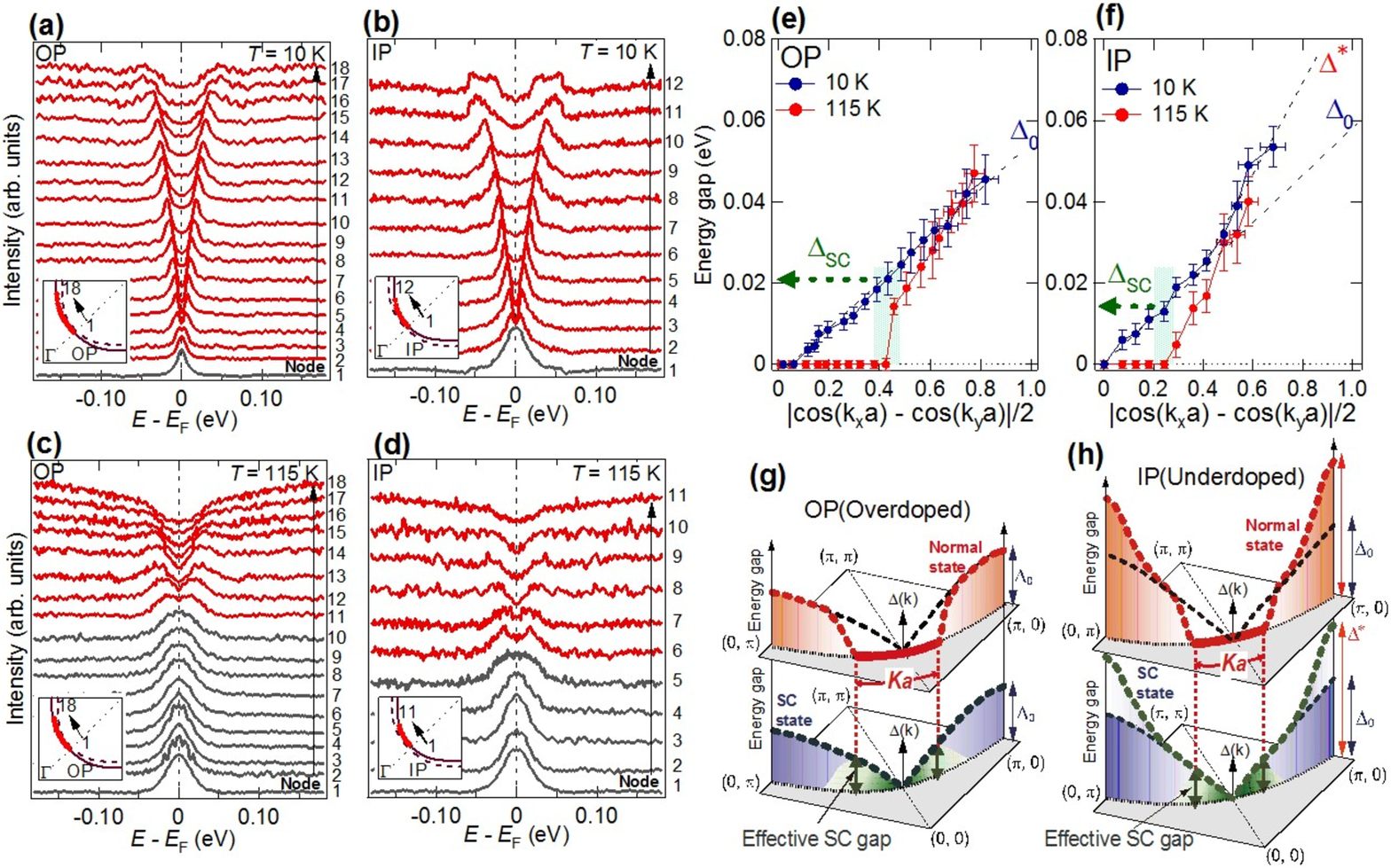}
\caption{(Color online) Temperature dependences of the energy gap for OP and IP. (a - d): Symmetrized energy distribution curves (EDCs) at Fermi momenta $k_{\rm{F}}$'s for the OP and IP bands in the SC ($T$ = 10 K) and normal ($T$ = 115 K) states. $k_{\rm{F}}$'s have been determined by the minimum gap locus \cite{M. R. Norman}. (e), (f): The values of the energy gaps shown in (a)$-$(d) plotted as functions of the $d$-wave order parameter $|\cos(k_xa)-\cos(k_ya)|$/2. Slightly above $T_c$ (= 115 K), the gap closes only in the nodal region (curves 1 - 10 for OP and 1 - 5 for IP), while it remains open in the anti-nodal region (curves 11 - 18 for the OP and 6 - 11 for the IP) \cite{1}. The effective SC gap $\Delta_{\rm{sc}}$ is defined by a gap at the edge of the Fermi arc. (g), (h): Schematic figures of the energy gap for OP and IP. Energy gap near the node $\Delta_0$, anti-nodal gap $\Delta^*$, Fermi arc length $K_a$ and effective SC gap $\Delta_{\rm{sc}}$ are shown.}
\end{figure*}
\begin{figure*}[ht]
\includegraphics[width=17.3cm]{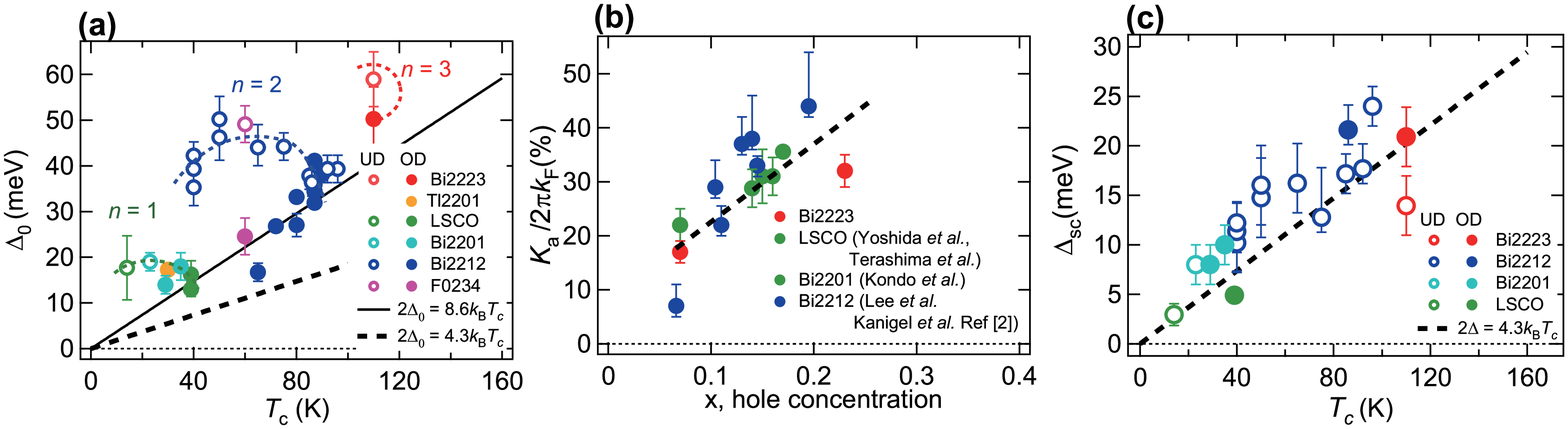}
\caption{(Color online) Relationship between the nodal SC gap $\Delta_0$, the Fermi arc length $K_a$, the effective SC gap $\Delta_{\rm{sc}}$ and $T_c$ for various HTSCs. (a): $\Delta_0$ as a function of $T_c$. Data are taken from ARPES results for LSCO \cite{Yoshida, Terashima}, Bi2201 \cite{Kondo}, Bi2212 \cite{Lee, Tanaka, Feng2, Ding}, and Ba$_2$Ca$_3$Cu$_4$O$_8$F$_2$ (F0234) \cite{Chen1}. These nodal gaps $\Delta_0$ have been estimated by linear fits of the data around the node. Arrows indicate the increase of hole concentration for each of $n$ = 1, 2, and 3. (b): $K_a$ relative to that of the full Fermi surface 2$\pi k_{\rm{F}}$ as a function of hole concentrations. Dashed line is a guide to the eye. (c): Effective SC gap $\Delta_{\rm{sc}}$ as a function of $T_c$.  The dashed line indicates $2\Delta_{\rm{BCS}}$ $\simeq$ 4.3 $k_{\rm{B}}T_c$, namely, the weak coupling $d$-wave BCS relationship.}
\end{figure*}
\begin{figure}[http]
\includegraphics[width=8.5cm]{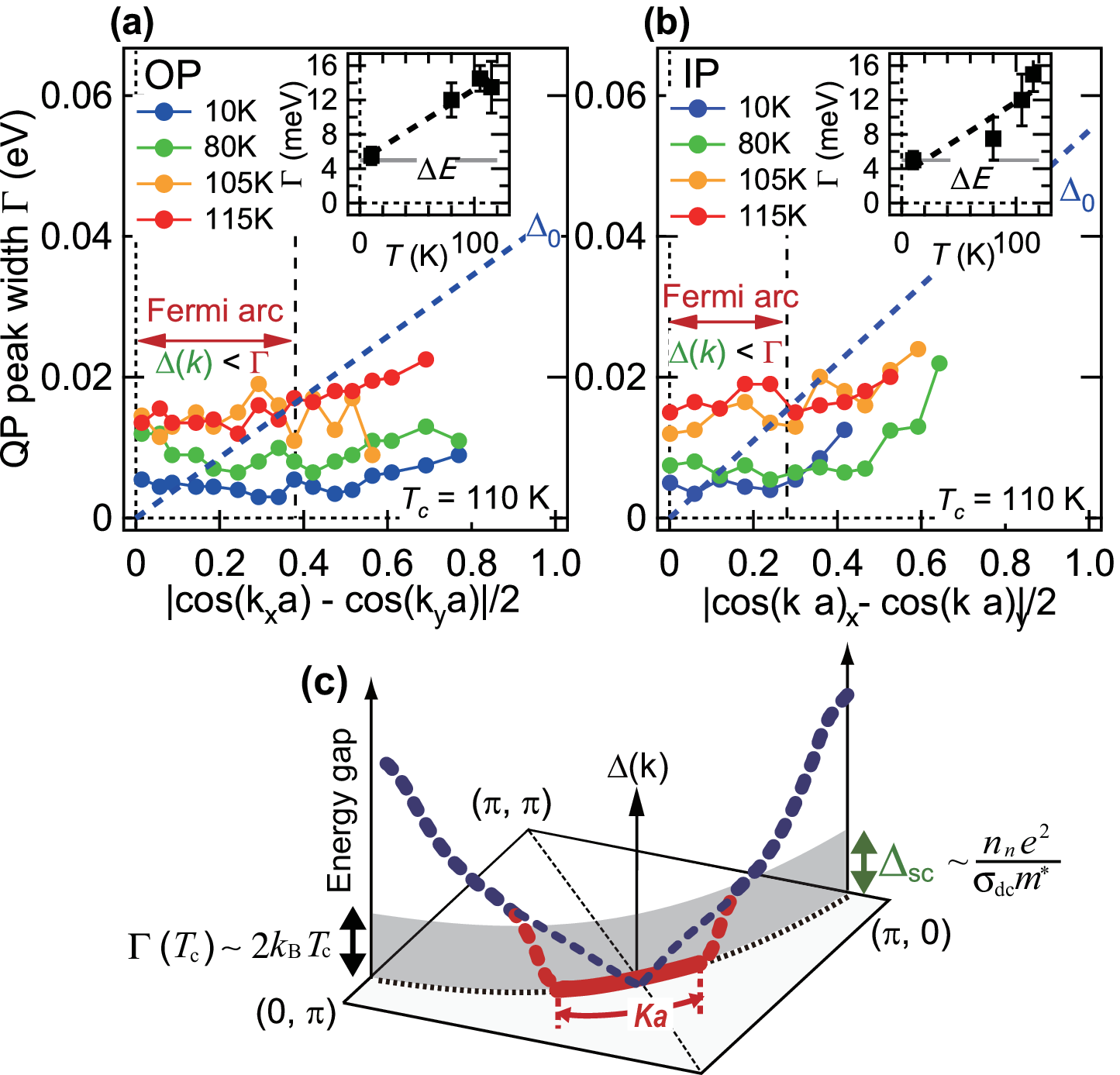}
\caption{(Color online) Temperature dependence of the quasi-particle (QP) width $\Gamma$ (HWHM). QP scattering rate $\Gamma$ at $k_{\rm{F}}$'s of OP and IP is plotted as a function of $d$-wave order parameter [(a): OP, (b): IP]. Blue dashed lines in (a) and (b) indicate the nodal SC gap $\Delta_0$. The Fermi arc appears in the momentum region where the gap magnitude $\Delta(k) \leq \Gamma$ is safisfied slightly above $T_c$, 115 K. The crossing point of the SC gap and $\Gamma$(115K) yields the end point of the Fermi arc and $\Delta_{\rm{sc}}$. Inset shows the temperature dependence of $\Gamma$ in the nodal region.  The experimental energy resolution $\Delta E$ is indicated. (c): Schematic figure showing the relationship between  $K_a$, $\Gamma$, $\sigma_{\rm{dc}}$ and $T_c$ for HTSCs in the SC and normal states. Blue dashed line shows the energy gap in the SC state, and red solid and dotted curves indicate the Fermi arc formation just above $T_c$. Gray region shows the energy which falls width $\Gamma$($T_c$) of QP at $T_c$. The energy gap within the gray area becomes zero around the node ($\Delta_{\rm{sc}}<\Gamma$) in the normal state. 
}
\end{figure} 

We show symmetrized energy distribution curves (EDCs) at Fermi momenta ($k_{\rm{F}}$) for OP and IP in Fig. 2. In the SC state [panels (a) and (b)], one can clearly see a monotonic increase of the gap from the nodal to the anti-nodal regions. In the normal state [panels (c) and (d)], the gap vanishes around the node, while in the off-nodal region, the gap remains open. In Figs. 2(e) and 2(f), the gap magnitudes for OP and IP are plotted as functions of the $d$-wave order parameter. The gap in the SC state shows a deviation from the ideal $d$-wave gap in the anti-nodal region of IP, namely, showing \textquotedblleft two-gap behavior\textquotedblright \cite{YRZ, Lee, Tanaka, Yoshida, Pushp}, and is extrapolated to $\Delta^* \sim$ 85 meV in the anti-nodal region as indicated in Fig. 2(f). On the other hand, the gap in the SC state for OP shows almost the ideal $d$-wave angular dependence, namely, \textquotedblleft one-gap behavior\textquotedblright as shown in Fig. 2(e). These \textquotedblleft two-gap\textquotedblright $versus$ \textquotedblleft one-gap\textquotedblright\ behaviors for IP and OP are consistent with the hole concentrations of IP (underdoped) and OP (overdoped), respectively, estimated from the ARPES experiment \cite{Trokiner}. Following the method employed in the previous ARPES studies \cite{Lee, Tanaka, Yoshida}, the IP gap data were fitted to two lines as shown in Fig. 2(f), while the OP gap data were fitted to one line. If we define the \textquotedblleft nodal gap\textquotedblright $\Delta_0$ by extrapolating the fitted line near the node to the anti-node direction, the $\Delta_0$ $\sim$ 60 meV for IP and $\Delta_0 \sim$ 50 meV for OP [Figs. 2(e) and 2(f)]. 

Slightly above $T_c$ ($T$ = 115 K), the Fermi arc appears around the node as shown in Figs. 2(e) and 2(f). The Fermi arc length $K_a$ defined by the FS region where the gap collapses in the symmetrized EDC are $\sim$ 32, and $\sim$ 17 $\%$ of the full FS length 2$\pi k_{\rm{F}}$ for OP and IP, respectively. If we define an effective SC gap $\Delta_{\rm{sc}}$ by the gap value at the edge of the Fermi arc, $\Delta_{\rm{sc}} \sim$ 21 $\pm$ 3 and $\sim$ 14 $\pm$ 3 meV for the OP and IP, respectively [Fig. 2(e) and 2(f)] \cite{Pasupathy}. 

Now, we discuss the central issue of which single parameter is related with the $T_c$ value of the HTSCs most directly. The well known Uemura relationship relates the superfluid density with $T_c$ \cite {Uemura1} for a wide range of HTSC's except for overdoped ones. Tallon $et\ al$. have proposed a modified Uemura relation\cite{Tallon} in which the value of $T_c$/$\Delta^{\prime}$ plotted as a function of superfluid density, where $\Delta^{\prime}$ is the maximum spectral gap obtained from the specific-heat and Raman studies. In Fig. 3(a), we have plotted the nodal SC gap $\Delta_0$ $versus$ $T_c$ for various HTSCs measured by ARPES, where the nodal gap $\Delta_0$ has been estimated by a linear fit of the gap data around the node as in the previous ARPES studies. For optimum to overdoped samples, the experimental data follow the linear relationship 2$\Delta_0$ $\sim$ 8.6$k_{\rm{B}}T_{c}$, reminiscent of a strong-coupling BCS formula of $d$-wave SC, whereas in the underdoped region, the plot becomes 2$\Delta_0 \gg 8.6k_{\rm{B}}T_c$, deviating from the linear BCS relationship. 

Then we take into account effects of the finite Fermi arc lengths $K_a$ as follows. In Fig. 3(b), $K_a$ for the OP and IP of Bi2223 as well as those for LSCO \cite{Yoshida, Terashima}, Bi$_2$Sr$_2$CuO$_{6+\delta}$ (Bi2201) \cite{Kondo}, and Bi2212 \cite{Lee, Kanigel} defined in the same way are plotted as a function of doped hole concentration $x$. One finds that the $K_a$ values of the various HTSCs approximately fall on a single line. Therefore, the evolution of $K_a$ with $x$ means the increase of the hole concentration in the CuO$_2$ plane in the normal state $n_n$($\approx x/a^2)$, where $a$ is the in-plane lattice constant, and hence is proportional to the superfluid density $\rho_s=n_s/m^{*}$ \cite{Kondo}, where $m^{*}$ is the carrier effective mass. Tanner $et\ al.$ \cite{Tanner} pointed out that the number of SC electrons $n_s$ is empirically proportional to $n_n$ ($n_s\sim0.2n_n$).

In Fig. 3(c), we have plotted the effective SC gap $\Delta_{\rm{sc}}$ which is approximately proportional to $K_a\Delta_0$ against $T_c$ for various HTSCs. These data approximately fall on the straight dotted line which represents the $d$-wave BCS gap 2$\Delta_{\rm{BCS}}$ = 4.3$k_{\rm{B}}T_c$ \cite{Won}. It should be noted that this relationship seems to hold even in the overdoped region. Hence, the \textquotedblleft effective\textquotedblright \ SC gap $\Delta_{\rm{sc}}$ (as indicated in Fig. 2) for all the HTSCs known to the authors is more directly related to $T_c$ than $\Delta_0$. This relationship is reminiscent of the relationship $T_c$ $\propto$ $x\Delta_0$ proposed by Lee and Wen \cite{LeePA} and Oda $et\ al.$ \cite{Oda}. The closer relationship between $\Delta_{\rm{sc}}$ and $T_c$ than that between $\Delta_0$ and $T_c$ is reasonable because the gap outside the Fermi arc does not close just at $T_c$ and hence has only small contribution to the condensation energy of superfluid.

While $\Delta_{\rm{sc}}$ satisfies the weak-coupling BCS formula 2$\Delta_{\rm{sc}}$ $\simeq$ 4.3$k_{\rm{B}}T_c$, $\Delta_0$ satisfies the strong-coupling relationship 2$\Delta_0 \simeq 8k_{\rm{B}}T_c$ as shown in Fig. 3(c) from the optimum to the overdoped regions such as the overdoped OP of Bi2223, where the second gap of magnitude $\Delta^*$, most likely due to competing order, does not exists in the anti-nodal region. This means that the strong-coupling $d$-wave superconductivity leads to the pseudogap beyond the Fermi arc region and the gap does not close just above $T_c$ there, even if the gap originates from superconductivity and not from the competing order.

In the remaining part of this letter, we address the question of why the Fermi arc length $K_a \propto x$ in the normal state. This is a key question because $T_c$ seems to be determined by $T_c \propto \Delta_{\rm{sc}} \sim K_a\Delta_0$, namely, by the paring strength $\propto$ $\Delta_0$ and the carrier/superfluid density $\propto K_a$. In order to answer this question, it should be remembered that the Fermi arc is a phenomenon which occurs at finite temperatures $\geq T_c$. Since finite temperatures broaden the quasi-particle width (QPW), we examine the temperature dependence of QPW measured by ARPES. In Fig.4, we have plotted the QPW at various temperatures and compare it with the $d$-wave gap. One can see that the QPW (HWHM) denoted as $\Gamma$ increases with temperature. An early ARPES study \cite{Valla3} has shown that the QPW around the node in the normal state is proportional to temperature $T$ as a fingerprint of marginal Fermi liquid \cite{Varma}. Within the experimental resolution of ARPES, $\Gamma$ $versus$ $T$ is consistent with $\Gamma \propto T$. From the present data at $T = 115$ K, QPW estimated at $k_{\rm{F}}$ just above $T_c$ is found to be $\Gamma \sim$ 20 meV ($\sim$ 2$k_{\rm{B}}T_c$). In the $k_{\rm{F}}$ region where $\Gamma$ is larger than the energy gap $\Delta(k)$ and hence  $\Gamma(\sim 2k_{\rm{B}}T_c) \geq \Delta(k)$, the gap cannot survive above $T_c$ and collapses into the Fermi arc. Therefore, the gap at the end point of the Fermi arc $\Delta_{\rm{sc}}$ should be approximately equal to $\Gamma \sim 2k_{\rm{B}}T_c$ slightly above $T_c$ and thereby one can re-derive the relationship $T_c \propto K_a\Delta_0$. This scenario is schematically shown in Fig. 4(c).

It is interesting to point out that the relationship $T_c \propto \Delta_{\rm{sc}} \sim K_a\Delta_0$ is consistent with Homes' law $n_s/m^* \propto \sigma_{\rm{dc}}(T_c)T_c$ \cite{Homes}, where $\sigma_{\rm{dc}}$ is the in-plane dc conductivity, as follows. If the effective SC gap $\Delta_{\rm{sc}}$ at the edge of the Fermi arc is determined by $\Delta_{\rm{sc}} \sim \Gamma(T_c)$ and if $\Gamma \sim 2k_{\rm{B}}T$ as mentioned above, one obtains 2$\Delta_{\rm{sc}} \sim 4k_{\rm{B}}T_c$. Therefore, since the in-plane dc conductivity slightly above $T_c$ is expressed as $\sigma_{\rm{dc}} \sim n_ne^2/m^*\Gamma(T_c)$ with $\Gamma(T_c) \sim 2k_{\rm{B}}T_c$, one obtains $T_c \sim k_{\rm{B}}n_ne^2/2m^*\sigma_{\rm{dc}}(T_c)$. Applying Tanner's law $n_s \sim 0.2n_n$ to the latter formula, from the ARPES point of view, too, $T_c \sim 5k_{\rm{B}}n_se^2/2m^*\sigma_{\rm{dc}}(T_c)$, Homes' law.  This means that the $T_c$ of HTSCs is dominated by the carrier density in the nodal region and the scattering rate of the carriers as pointed out by Zaanen in the context of Planckian dissipation \cite{Zaanen}.

We thank C. Panagopoulos for informative discussion. ARPES experiments were carried out at HiSOR, Hiroshima Synchrotron Radiation Center, Hiroshima University (Proposal No. 07-A-10, and 08-A-35). This work was supported by Grant-in-Aid for Young Scientists (B), and by a Global COE Program \textquotedblleft the Physical Sciences Frontier\textquotedblright, MEXT, Japan. SI acknowledges support from the Japan Society for the Promotion of Science for Young Scientists.\\


\begin{thebibliography}{99}
\bibitem{1}The magnitude of the gap has been determined by the peak positions of the symmetrized EDCs in the SC state. For the off-nodal spectra in the normal state, there is no clear peak. Therefore, the position of the peak (hump) is determined by the second derivative of the spectra, and is used to estimate the gap size.
\end{thebibliography}
\end{document}